# A Fault-tolerant Structure for Reliable Multi-core Systems Based on Hardware-Software Co-design

Bingbing Xia, Fei Qiao, Huazhong Yang, and Hui Wang
Institute of Circuits and Systems, Dept. of Electronic Engineering
E-mail:xbb07@mails.tinghua.edu.cn, {qiaofei, yanghz, wangh}@tsinghua.edu.cn

## Abstract

To cope with the soft errors and make full use of the multi-core system, this paper gives an efficient fault-tolerant hardware and software co-designed architecture for multi-core systems. And with a not large number of test patterns, it will use less than 33% hardware resources compared with the traditional hardware redundancy (TMR) and it will take less than 50% time compared with the traditional software redundancy (time redundant).Therefore, it will be a good choice for the fault-tolerant architecture for the future high-reliable multi-core systems.

## 1. Introduction

With the development of the semiconductor technology, integrating billions of transistors on a single die has been possible and the complexity of the SoC is increasing. Further, more and more processors and IP cores can be implemented on a single die to build a multiprocessor SoC. Along with this trend, the electronic devices are becoming more and more sensitive to external disturbs such as soft errors [1], and fault-tolerant architecture is always used to obtain the reliability.

For fault-tolerant architectures, hardware redundancy and software redundancy are the two popular methods and each kind can be classified further according to whether it's static or dynamic. FTMR (Coming from the N-Modular Redundancy method) is the popular static hardware redundancy method while N-version programming is the typical static software redundancy [2][3][4]. During these methods, no matter static or dynamic, software redundancy has more flexibility while the hardware redundancy has less hardware resources needed. The proposed architecture in this paper takes a combination of these two methods to make full use of the multi-core system to realize the fault-tolerancy with both the timing and resources constraints taken into consideration.

Laterly, George A.Reis gives a software implemented fault tolerance mechanism named SWIFT with a enhanced control-flow checking mechanism based on the compiler technology[5].This mechanism is useful for software fault -tolerant, but do nothing with the other related hardware modules. S.Tosun gives a reliability-centric hardware/software co-design framework to partition the hardware/software which takes reliability into consideration which is useful for the initial task scheduling phase [6]. This framework is useful for the initial hardware/software partitioning, but it can't tolerate the transient faults such as soft errors dynamically. These two methods belong to static methods for the reliable system design. In addition, a combined software and hardware technique for the design of reliable IP processor is given by M.Rebaudengo, but it puts software redundancy at the first place and uses partial hardware redundancy, the aim is designing a reliable single-core processor [7].It doesn't consider making full use of the hardware and software resources for the multi-core systems.

The presented architecture is different from the methods listed above, which is trade-off between software redundancy and the hardware redundancy, and can dynamically tolerate the fault based on the BIST (Built-In-Self-Test) structure. Moreover, the new method needs less hardware resources than the hardware redundancy, while which has the flexibility like the software one. The basic idea is using software to replace the hardware module once the fault is detected by the BIST structure, and it's especially suitable for the multi-core systems today. In multi-core systems, as limited by the algorithm's parallelism, not all the cores are busy all the time. So these spare processor cores can be made full use of to realize the fault-tolerant with the architecture proposed here.

The contribution of this paper is as follows:

Firstly, a new architecture is come up which makes use of hardware/software co-design [8] and BIST[9][10] technology to realize the fault-tolerant system. Secondly, a methodology based on this architecture is given for time-limited fault-tolerant system designs. Thirdly, experiments are implemented to show the advantages of the architecture and an application of QoS scalable MPEG2 video decoder is given to overcome the disadvantage of this architecture for its usage in real-time multimedia designs.

The next section gives an overview of the architecture for this fault-tolerant system model. The design methodology based on such a fault-tolerant architecture is illustrated in detail in section 3. Section 4 shows the experimental results and gives an application of QoS scalable MPEG2 video decoder to show how to overcome the disadvantage in real-time multimedia designs. Finally, section 5 draws the conclusion and gives the future work related.

## 2. Descriptions of the architecture

The proposal is targeted at multi-core SoC which is made up of many processors cores and hardware modules. In this fault-tolerant architecture, hardware-software co-design technology is used to exploit the fault-tolerant metric, which is a trade-off between software flexibility and hardware high-efficiency.

The system architecture for multi-core SoC is shown in figure 1.

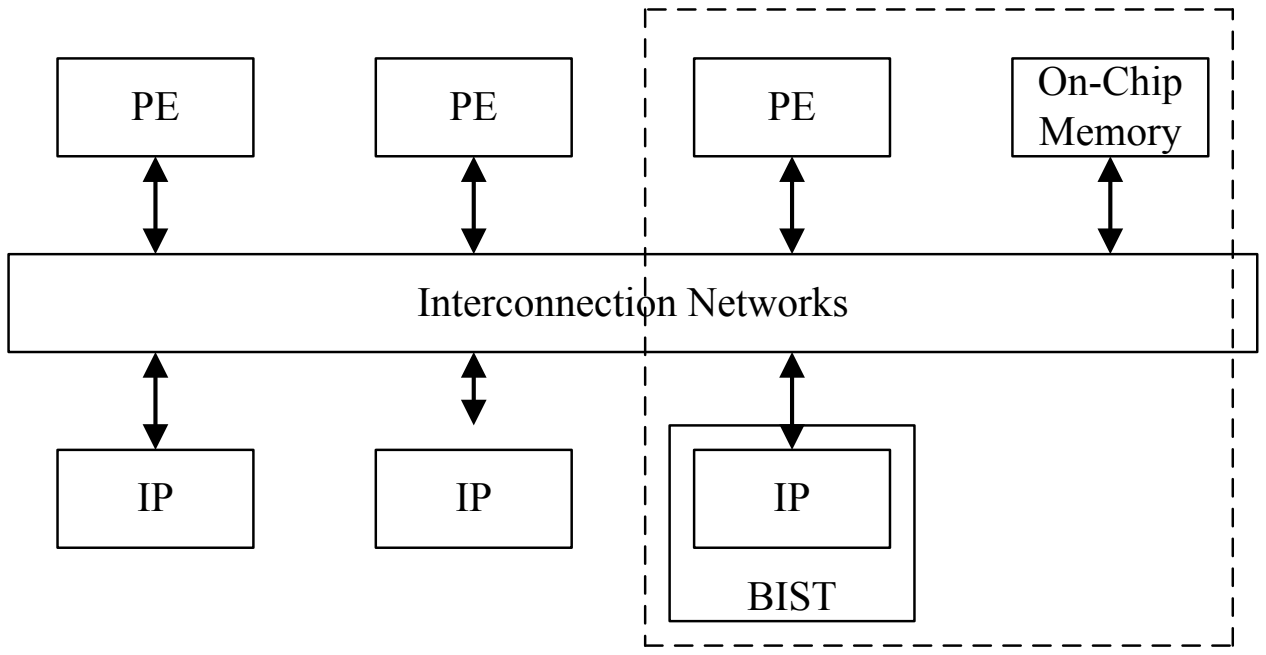

**Figure 1**: Fault-tolerant Multi-core System Architecture

In this architecture, the multi-core SoC is made up of PEs (Processing Element, in this paper , it refers to processor core ), hardware IP modules and the interconnection network that connects them together. In this system, some IP modules are selected to be fault-tolerant and they are used in the proposed fault-tolerant architecture, such as the parts located in the dashed border. The detailed illustration of these parts is in figure 2.

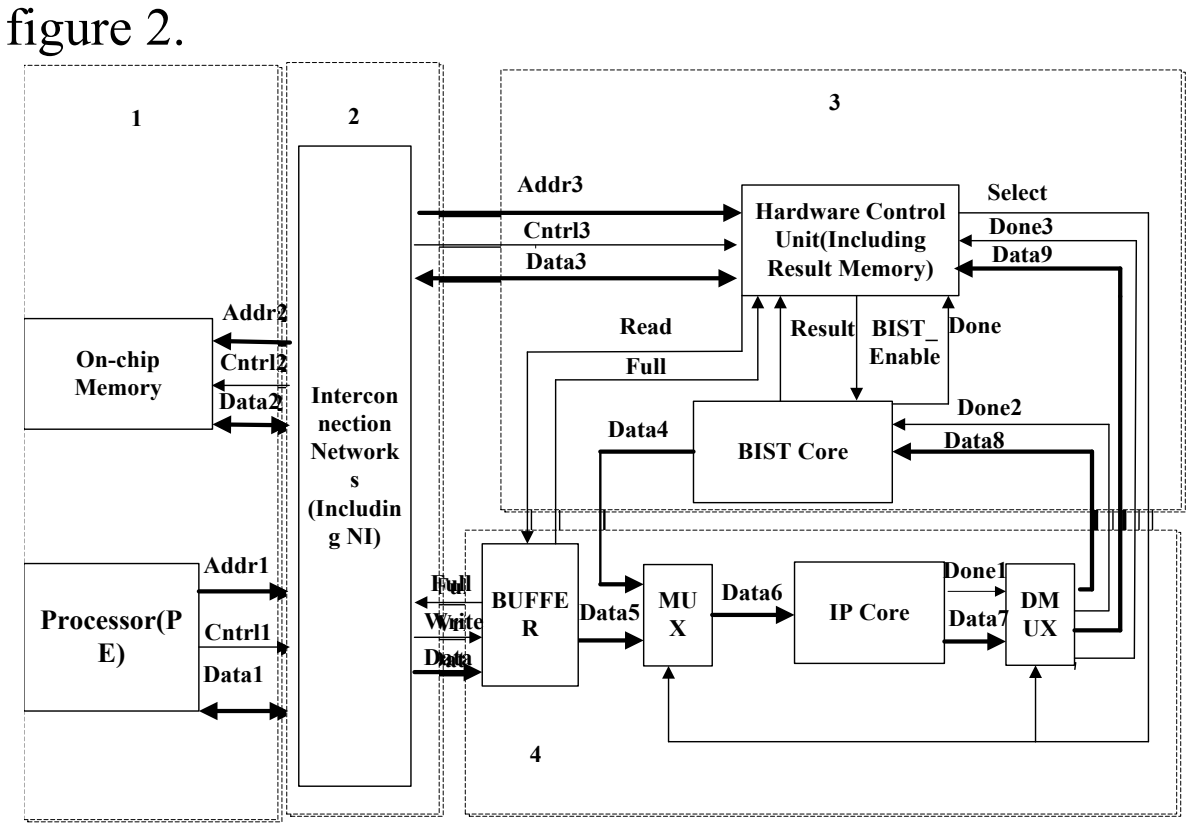

**Figure 2**: Fault-tolerant architecture for the particular IP module

This architecture is made up of three main parts except the on-chip memory, one is the processor (processors), the other is the BIST test structure, and the third is the interconnection network that connects them together. In this architecture, the IP core is put under test and acts as the hardware accelerator. The hardware control unit is used to control all the parts of the BIST structure and supplies the interfaces to the interconnection networks to acquire the test patterns and the related correct results for comparison. The BIST Core is made of three components: the TPG (Test Pattern Generator) which transfers the test patterns from the hardware control unit, the TRA (Test Response Analyzer) which is used to analyze the result and the BIST control unit for the BIST control procedure. The BIST test structure receives the test patterns and the correct results from the memory through the interconnection networks (In DMA (Direct Memory Access) Mode), and once the test is over , the result will be returned to the processor to tell it whether there is fault in the IP core. The test patterns are stored based on a priority-based method which store and select the test patterns in such a way: the test pattern that relates to the most sensitive fault or the one that we care most will be put in the first place and gives the highest priority, and then choose the second one, so on and so forth. During the test procedure, run as more test patterns as possible to get a better fault coverage. After the BIST test, the processor will take response according to this result.

The total flow of designing such a fault-tolerant architecture is as follows:

Firstly, partition the hardware and software based on a particular multi-core architecture and parallelize the redundant software code which will act as the function of the hardware module once hardware fault is detected.

Secondly, through the system simulation, an initial estimation of the number of the clock cycles that will be taken to test the module is obtained. And based on this number, modify the machine code on the processor to insert a request signal to the BIST test structure to start the BIST function.

Thirdly, BIST begins and DMA transfer is used to transfer the test patterns and the correct results to the BIST test structure from the memory. While at the same time, the software is still running until when the function taken by the hardware module begins. Then the processor arrives at a break and waits for the result of BIST result.

Fourthly, BIST test ends and the BIST test structure gives an ack signal to the processor and gives a 6ne bit signal to show whether there is fault or not.

Finally, the processor will decide whether use the hardware module or software to realize the function. If software function is used, spare cores will be used to accelerate the software to make full use of the multi-core system. While hardware module is fault-free, the function will be implemented on the hardware module to accelerate the function.

The advantages 6f this architecture are as follows:

From the view of the multi-core system design, it makes full use of the idle cores in the multiprocessor system to exploit the fault-tolerant mechanism, in such systems, some processor (like ASIP (Application Specific Instruction Processor) is selected for a specific application, such as the multimedia usage. And in such circumstance, the software running time may be not much more than the IP core and may also meet the need for the real-time applications. In this way, this architecture can be used to efficiently reduce the hardware cost introduced by the hardware-redundancy such as the FTMR.

The most important advantage of this architecture is, compared with the hardware-redundant method, it saves the hardware resources, while with comparison to the time-redundant software-redundant method, it is more time-efficient supposing that the hardware fault rate will be very low. What's more, as the test patterns are transferred using the DMA method from the outside memory, this method will use less hardware than the ordinary defect-oriented BIST structure which stores the specified test patterns with the on-chip memory[11].

Taking the reliability of the whole digital system into consideration. According to the formula, the fault-free probability 6f a single transistor changing with time is as follows [12]:

$$P = e^{-(\lambda_1 + \lambda_2)t}$$

(1)

Here, P is the fault-free probability, t is the time and $\lambda_1 + \lambda_2$ is about $10^{-5} h^{-1}$, suppose that each transistor contributes equally to the total system ,then the whole reliability (fault-free probability) of the system with N transistors is:

$$P_{whole} = e^{-N*(\lambda_1+\lambda_2)t}$$

(2)

Based on this formula, the fault-free probability of our architecture is as follows:

$$P_{whole} = e^{-4*10^{-5}*numberofgates(NAND)*t}$$

(3)

In the formula above, the number of the MOS transistors is evaluated by the number of the NAND gates which uses 4 MOS transistors each. Based on this , the proposed architecture needs less hardware resources than hardware-redundancy and needs less memory space than the ordinary software redundancy (say, N-version programming or the time-redundant), so it will be more reliable this advantage will be more significant with the increase of the hardware module scale or the software one.

## 3. The Proposed Design Methodology Based on Such A Fault-Tolerant Architecture

For this architecture, scheduling of tasks (both hardware and software tasks) is very important, especially for the real-time applications [13]. Since this architecture is especially useful for heterogeneous multi-core systems, it should be taken into consideration at the system-level design step and give an estimation of the cost to determine whether to use this architecture and how to use it.

Suppose that the memory space of the processor is enough for all the software codes. And since the test structure (apart from the test pattern and test result pattern memory) is very simple compared with the complex IP core under test, the hardware resources taken by it can be ignored. What's more, since the communication time can be overlapped through the DMA transfer and data-prefetch [14], its impact on the whole system is much less than the other parts.

Total runtime for the hardware-redundancy (here, using FTMR as the method) is:

$$T_{hr} = T_{s1} + T_h + C_c$$

(4)

In equation (4), $T_{hr}$ is the total runtime for the hardware-redundancy, $T_{s1}$ is the software running time for the whole algorithm without the function that uses hardware to accelerate. $T_h$ is the hardware running time for the function which uses the hardware accelerator. $C_c$ is the communication cost between the processor and the hardware accelerators.

T6tal hardware resource for the hardware-redundancy mechanism is:

$$H_{hr} = H_p + 3 \times H_h + M_{s2}$$

(5)

In equation (5), $H_{hr}$ is the total hardware resource for the hardware-redundancy mechanism, $H_p$ is the hardware resource used for the processor, $H_h$ is the hardware resource used for the hardware module under test. $M_{s2}$ is the memory space used for the storage of the software code whose function equalizes the hardware module.

Total runtime for the software-redundancy (here, using 3 version programming as the method and three versions are running in parallel) is:

$$T_{sr} = T_{s1} + T_{sf}$$

:

(6)

In equation (6), $T_{sr}$ is the total runtime for the software-redundancy, $T_{s1}$ is the same as the one in formula 3, $T_{sf}$ is the software running time for the function that can be accelerated by the hardware.

Total hardware resource for the software-redundancy is:

$$H_{sr} = 3 \times H_p + 3 \times M_{s2} + M_{s1}$$

(7)

In equation (7), $H_{sr}$ is the total hardware resource for the software-redundancy. $H_p$ and $M_{s2}$ are the same as those in equation (4) , $M_{s1}$ is the memory space used for the storage of the software code without the part that will be accelerated by the hardware module.

Total runtime for the proposed architecture is:

$$T_{pr} = T_{s1} + T_{hf} \times (1 - P_{fault}) + T_{sf} \times P_{fault}$$

(8)

In equation (8), $T_{pr}$ is the total running time for the proposed architecture. $T_{s1}$ and $T_{sf}$ are same as those in equation (5) . $P_{fault}$ is the probability that the hardware meets a fault. $T_{hf}$ is the hardware running time for the function which is using the hardware accelerator.

Total hardware resource for the proposed architecture is:

$$H_{pr} = H_p + M_{s1} + M_{s2} + H_h + M_{test-pattern} + (N-1) \times M_{test-pattern}$$

(9)

In equation (9), $H_{pr}$ is the total hardware resource for the proposed architecture, $H_p$, $M_{s1}$, $M_{s2}$ and $H_h$ are the same as those listed above , $M_{test-pattern}$ is the memory space used by each test-pattern , and N is the number of test patterns for BIST procedure. Here, since $M_{s1}$, $M_{s2}$ and $(N-1)\times M_{test-pattern}$ are always on the RAM that can be put outside the chip core , so they can be ignored. And for the selection of the number N, it should be a good trade-off between the test-time and the fault coverage, for a particular application, only a small part of the whole test patterns should be mostly cared about, so as less test patterns as possible should be used to get the initial test result.

The total design flow for such fault-tolerant multi-core systems is shown in figure 3. At the beginning, an initial hardware/software partitioning is acquired based on the simulation for the processors and the hardware modules. And the minimum number of test-patterns for each fault-tolerant hardware module can be obtained according to the simulation. Then, the constraints of each fault-tolerant hardware module can be obtained, here, the most 2 important constraints are the hardware resource limit and the running deadline constraint for each one.

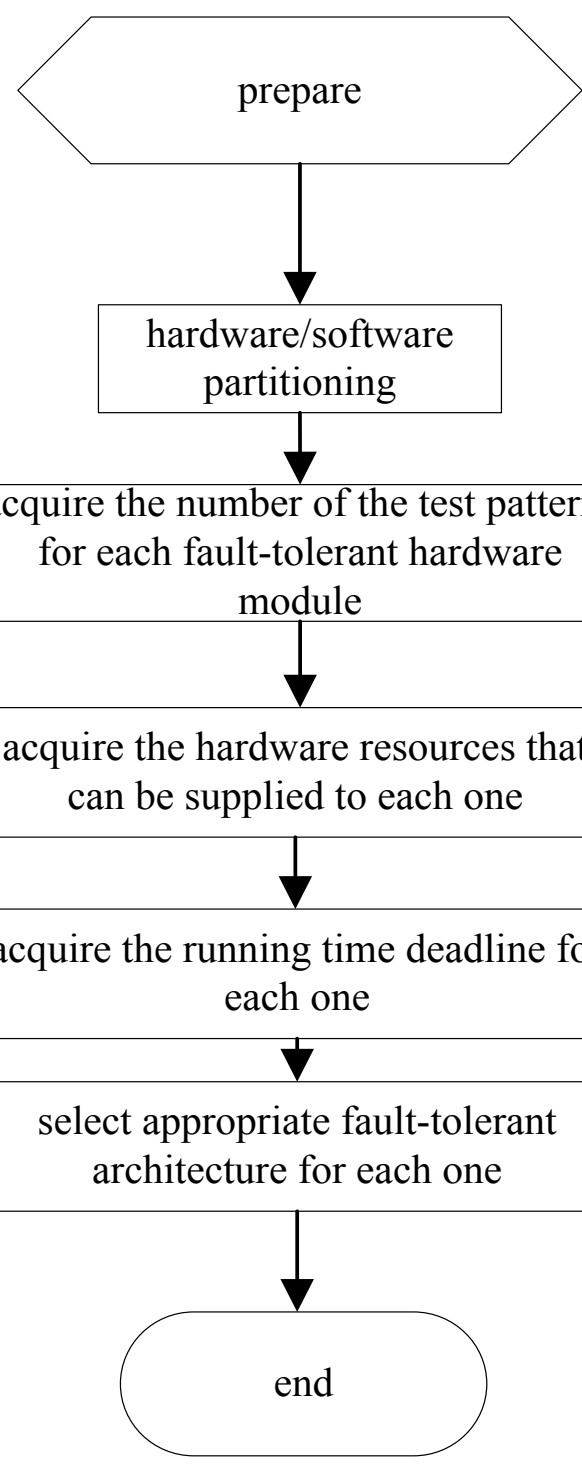

**Figure 3**: design methodology for fault-tolerant multi-core systems

Based on this partitioning, select the appropriate fault-tolerant architecture for each of them in the following way:

Supposing the on-chip hardware-resources that can be supplied to this hardware module is HT, the running time deadline that can be taken by this hardware module is TT. Then:

IF HT > $H_p + 3\times H_h$

Use the FTMR method

ELSE

IF HT > $H_p + H_h + M_{test-pattern}$

Use the proposed architecture

ELSE

IF TT > $T_{s1} + T_{sf}$ AND HT > $3\times H_p$

Use the 3-version programming time-redundant software-redundant architecture

ELSE

Try other software-redundant architectures (10)

END IF

END IF

END IF

Note that in (10), if other software-redundant methods (such as recovery block and 2-version programming, etc) can't meet the constraints, the partitioning should be considered twice and refresh it to get a appropriate portioning for such a fault-tolerant system.

After selecting the fault-tolerant architecture for each hardware module is finished, then the total methodology ends and the system level fault-tolerant system is developed.

## 4. Experimental results

In this part, three kinds of experiments are implemented for the evaluation of the proposed architecture and an application of QoS scalable MPEG2 video decoder is shown to overcome the drawbacks of such fault-tolerant system for the real-time multi-media systems.

### 4.1. Resource usage evaluation and performance comparison experiments:

Two experiments are implemented, one for combinational circuit module and the other for sequential circuit module. A sorting module is made as the combinational circuit module and an IDCT module is taken as the sequential circuit module, the results are as follows:

Supposing the fault rate is 1%, then through the RTL design and use the Design Compiler, the results are as follows: (X is the hardware logic resources needed by the MIPS processor, about 40000 gates). Here, the processor is the MIPS processor which we have designed by ourselves. The test results include the simulation outcomes of both the processor and the hardware module. In these experiments, the software on the processor is for the initialization of the data prepared for the hardware module under test.

The methods chosen for the comparison are as follows:

Hardware redundancy uses the popular NMR method (here, using FTMR method which uses three copies of the module and the majority voter to get the right result). Software redundancy uses the popular N-version programming method (Here N = 3 and change little between different versions).

#### 4.1.1. Hardware logic and running time analysis result:

Table 1 shows the test result of the running time and hardware logic for the comparison of the three architectures for the case of a combinational circuit module (sorting module). From this table, it can be seen that the running time of the proposed architecture is little more than the hardware redundancy, but it needs much less hardware logic than the hardware redundancy .What's more the running time of the proposed architecture is much less than the software redundancy , while the hardware logic needed is more than the software redundancy. In a word, compare these three methods with a ratio (which is performance/logic number, here, using the 1/ (Running Time) as the performance value), it can be concluded that:

Hardware Redundancy : Software Redundancy : Proposed Architecture = 4.9026 : 2.6344 : 8.8299

This means the proposed architecture will get the best performance per each logic gate.

**Table 1:** Test results of the running time and hardware logic for the comparison of the three architectures for the case of a combinational circuit module (sorting module)(N=4)

| | Hardware Logic (unit:gate) | Running Time (unit:Cycle) |
|---|---|---|
| Hardware Redundancy | 111789+X | 4019 |
| Software Redundancy | 3X | 9490 |
| Proposed Architecture | X+43417 | 4073 |

Table 2 shows the test result of the running time and hardware logic for the comparison of the three architectures for the case of a sequential circuit module (IDCT module).From this table, same conclusions can be got as from table 1 while the ration comparison result is as follows:

Hardware Redundancy :Software Redundancy: Proposed Architecture =12.4209 : 3.7236 : 24.3144

This also shows that the proposed architecture will get the best performance per each logic gate.

**Table 2:** Test result of the running time and hardware logic for the comparison of the three architectures for the case of a sequential circuit module (IDCT module)(N=4*64)

| | Hardware Logic(unit:gate) | Running Time(unit:Cycle) |
|---|---|---|
| Hardware Redundancy | X+130692 | 1415 |
| Software Redundancy | 3X | 6714 |
| Proposed Architecture | X+44106 | 1467 |

### 4.1.2.Power Analysis Result:

Table 3 and Table 4 give the energy consumed by each of the fault-tolerant methods, table 3 is for the combinational circuit module while table 4 is for the sequential circuit module. From these results it can be concluded that the energy consumed by the proposed architecture is more than the hardware redundancy method (about 77%), but is much less than software redundancy method (the software redundancy energy is more than 36 times as large as the proposed architecture).

**Table 3:** power analysis result for the combinational circuit module

| | Energy(Power ×Running Time) | N |
|---|---|---|
| Hardware Redundancy | 0.01367mJ | --------- |
| Software Redundancy | 1.10835J | --------- |
| Proposed Architecture | 0.02387mJ | 4 |

**Table 4:** power analysis result for the sequential circuit module

| | Energy(Power ×Running Time) | N |
|---|---|---|
| Hardware Redundancy | 0.01393mJ | --------- |
| Software Redundancy | 0.73200mJ | --------- |
| Proposed Architecture | 0.02934mJ | 4*64 |

The main title (on the first page) should be at the top of the printable area, centered, and in Times 14-point, boldface type. Capitalize the first letters of nouns, pronouns, verbs, adjectives, and adverbs; do not capitalize articles, coordinate conjunctions, or prepositions (unless the title begins with such a word). Leave one blank line after the title.

### 4.2. Communication Cost And Optimization Method

Using the proposed architecture to implement the total system, both power and time cost are considered in the system architecture level, and to reduce the communication cost, both data-prefetch technology and DMA transfer method are used to get the transfer-time overlapped and power will be saved because the processor will no longer need to read the data and write them to the bus. Table 5 shows the comparison for the time and energy per each data fetch before and after the optimization (using 100MHz AMBA [15] bus as the transfer media, and using burst transfer mode). From this table, it can be seen that the DMA transfer overlaps the data fetch with the software running so the time can be ignored, while the energy per fetch can be reduced by nearly 39.7%. So these optimizations can be used to improve the system's performance which uses the proposed fault-tolerant architecture.

**Table 5:** The Time and Energy Each Fetch Before and After the optimization

| Module | Before Optimization | | After Optimization | |
|---|---|---|---|---|
| | Time | Energy | Time | Energy |
| IDCT | 128(cycles) | 1195.4nJ | 0(cycles) | 720.37nJ |
| Compare | 32(cycles) | 796.96nJ | 0(cycles) | 533.21nJ |

### 4.3. Fault-Injection And Coverage Test

Since this architecture mainly targets at the particular faults and the testing time is limited, so the coverage will be lower than typical BIST methods.

Table 6 shows the fault-coverage rate of the IDCT module. The fault-injection is on the gate level and 30 random faults are added to the model to simulate the test structure. From this table, it can be seen that the coverage rate is lower than the ordinary test structures. Therefore, this architecture can't ensure the full test coverage but can be used for the specified fault that is most sensitive to the environment.

**Table 6:** Coverage Rate Test Result of the IDCT module

| Number of blocks | Fault-Coverage |
|---|---|
| 1 | 63% |
| 2 | 83% |
| 3 | 87% |

### 4.4. Fault-free Probability Comparisons

According to equation (3) and the experiment results listed above, the fault-free probability comparisons can be easily acquired and the curves are as follows:

Fig 4 and fig 5 give the curves for the IDCT module and the sorting module, from these curves, conclusion can be drawn that the proposed architecture has the highest fault-free probability than the other 2 methods.

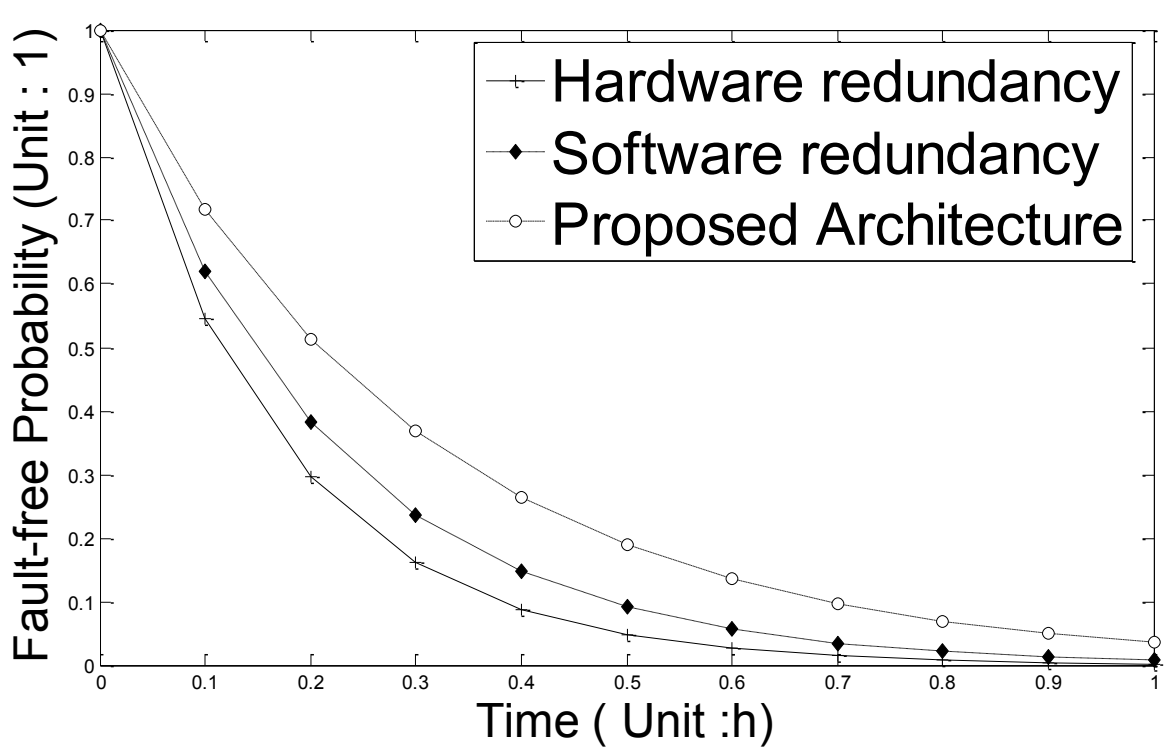

**Figure 4**: Fault Free probability curve for the IDCT module

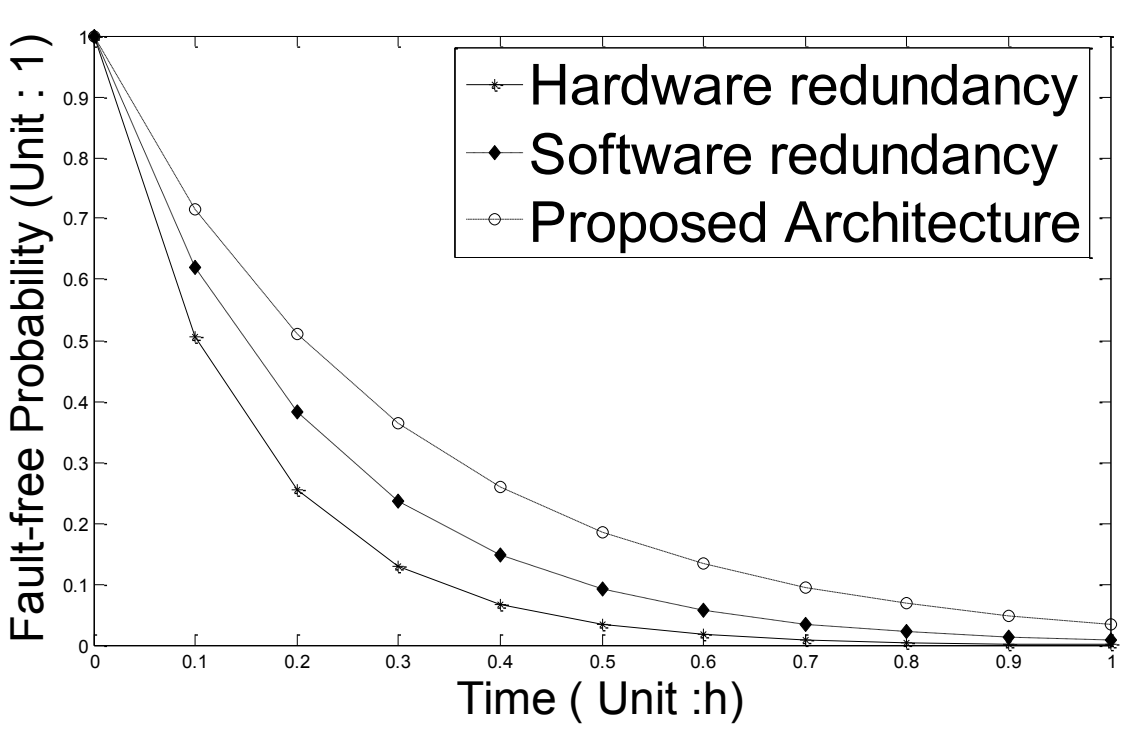

**Figure 5**: Fault Free probability curve for the sorting module

### 4.5. QoS Scalable MPEG2 video decoder with this proposed architecture

Since the software running time for the same function is always much more than the hardware accelerators, the proposed architecture can't ensure real-time when a fault occurs. To overcome this drawback, this part gives a QoS Scalable MPEG2 video decoder design for such real-time multimedia applications. MPEG2 [16][17] is a very popular video standard and is selected as the digital TV standard for most digital TVs in China. Once the hardware module of the decoder system has fault in it, it leads the system to a failure, there will be no picture on TV, and the whole system will be shipped to the factory for the repayment. Using the proposed architecture, many faults will be tolerated and the video can be continued in a low QoS mode instead of displaying no picture at all. Using SystemC [18][19][20] as the description language, such a MPEG2 decoder with the proposed architecture is implemented and simulated. In this architecture, IDCT module acts as the hardware module and the other functions of the decoder is put on the processor running in software(here ,using the powerful AMD Turion 64 processor as the main processor on which the software is running) . To make the results more reliable, standard test sequences are used to act as the input into the system.

The results are obtained as in table 7:

**Table 7:** Simulation Results for the QoS Scalable MPEG2 Video Decoder

| Frame Size | Test Sequences | Fault Existence | Average Decoding Cycles/Frame |
|---|---|---|---|
| 1920×1080 (HD) | riverbed | Yes | 23321460 |
| | | No | 21680195 |
| 1280×720 | Parkrun mobcal | Yes | 11269878 |
| | | No | 9881389 |
| 720 × 576 | parkrun mobcal | Yes | 4963147 |
| | | No | 4483583 |
| 352 × 288 (CIF) | mobile, foreman,football | Yes | 1345557 |
| | | No | 1166834 |
| 176 × 144 (QCIF) | mobile,fore man highway, news | Yes | 440084 |
| | | No | 396953 |

Based on table 7, suppose the processor and the hardware modules work at 550MHz, then the frame-rate for the HD video can be 25.4frames/second with no fault in the system but when a fault occurs, the frame-rate is lowered to 23.6frames/second.Thus, if a fault occurs in the hardware module (IDCT), the decoder can't realize the real-time decoder for the HD video. Using this fault-tolerant architecture, QoS can be lowered that decreases the frame-size to 1280×720 once a fault occurs and the frame-rate can be 48.8frames/second. This will be much better than no picture displayed at all in normal systems when a hardware failure occurs, although the QoS is somewhat lowered. This case analysis shows the how to overcome the disadvantage of this fault-tolerant architecture and its usage in multimedia applications.

## 5. Conclusions and future work

This paper gives an hardware-software co-design fault-tolerant architecture which is targeted at high-reliability multi-core systems. And through the analysis, conclusion can be drawn that the proposed architecture makes full use of the hardware and the software which are abundant in the heterogeneous multi-core systems. And if N is not large for the specific application it has less hardware resource needed than hardware redundancy and is faster than the normal software redundancy. The example of the MPEG2 QoS scalable video decoder shows the usage of this architecture for multimedia processing to overcome the drawback of the software running. Therefore, this architecture is very appropriate for the future heterogeneous multi-core systems which need high reliability. In our future work, this architecture will be used in heterogeneous multi-core system designs to acquire the high reliability, especially for the embedded systems. And it will benefit a lot for such designs. The disadvantage of the architecture is that the fault coverage is not very high compared with the normal BIST method, so in the future more attention will be paid to this limitation to improve the architecture.